\documentclass[acmtog,nonacm]{acmart} % add nonacm

\renewcommand\footnotetextcopyrightpermission[1]{} % removes permission footnote
\AtBeginDocument{%
  }

\usepackage{tcolorbox}
\usepackage{float} 
\usepackage[nameinlink,noabbrev]{cleveref}

\usepackage[disable]{todonotes} % or remove [disable] while drafting

\usepackage{soul}

\usepackage{siunitx}

\usepackage{subcaption}
\usepackage{multirow}
\usepackage{algorithm}
\usepackage{algpseudocode} % for \begin{algorithmic} ... \end{algorithmic}
\usepackage{listings}

\newcommand{\RQI}{How much does developer ownership-based clustering improve the performance of LLM-based LLPs?}
\newcommand{\RQII}{How much does semantic intent-based clustering improve the performance of LLM-based LLPs?}
\newcommand{\RQIII}{How much does combining semantic intent and ownership signals via multiplex clustering further enhance LLP performance?}

\begin{document}

\title{OmniLLP: Enhancing LLM-based Log Level Prediction with Context-Aware Retrieval}

\author{Youssef Esseddiq Ouatiti}
\affiliation{%
  \institution{Queen's University}
  \city{Kingston}
  \state{ON}
  \country{Canada}
}
\email{youssefesseddiq.ouatiti@queensu.ca}

\author{Mohammed Sayagh}
\affiliation{%
  \institution{ETS - Québec University}
  \city{Montreal}
  \state{QC}
  \country{Canada}}
\email{mohammed.sayagh@etsmtl.ca}

\author{Bram Adams}
\affiliation{%
  \institution{Queen's University}
  \city{Kingston}
  \state{ON}
  \country{Canada}
}
\email{bram.adams@queensu.ca}

\author{Ahmed E. Hassan}
\affiliation{%
  \institution{Queen's University}
  \city{Kingston}
  \state{ON}
  \country{Canada}
}
\email{ahmed@cs.queensu.ca}


\begin{abstract}
\textbf{Abstract. } Developers insert logging statements in source code to capture relevant runtime information essential for maintenance and debugging activities. Log level choice is an integral, yet tricky part of the logging activity as it controls log verbosity and therefore influences systems' observability and performance. Recent advances in ML-based log level prediction have leveraged large language models (LLMs) to propose log level predictors (LLPs) that demonstrated promising performance improvements (AUC between 0.64 and 0.8). Nevertheless, current LLM-based LLPs rely on randomly selected in-context examples, overlooking the structure and the diverse logging practices within modern software projects. In this paper, we propose OmniLLP, a novel LLP enhancement framework that clusters source files based on (1) semantic similarity reflecting the code's functional purpose, and (2) developer ownership cohesion. By retrieving in-context learning examples exclusively from these semantic and ownership aware clusters, we aim to provide more coherent prompts to LLPs leveraging LLMs, thereby improving their predictive accuracy. Our results show that both semantic and ownership-aware clusterings statistically significantly improve the accuracy (by up to 8\% AUC) of the evaluated LLM-based LLPs compared to random predictors (i.e., leveraging randomly selected in-context examples from the whole project). Additionally, our approach that combines the semantic and ownership signal for in-context prediction achieves an impressive 0.88 to 0.96 AUC across our evaluated projects. Our findings highlight the value of integrating software engineering-specific context, such as code semantic and developer ownership signals into LLM-LLPs, offering developers a more accurate, contextually-aware approach to logging and therefore, enhancing system maintainability and observability.
\end{abstract}

%    <concept>
%        <concept_id>10011007.10011074.10011111.10011696</concept_id>
%        <concept_desc>Software and its engineering~Maintaining software</concept_desc>
%        <concept_significance>500</concept_significance>
%        </concept>
%  </ccs2012>

% \ccsdesc[500]{Software and its engineering~Maintaining software}

\maketitle

\section{Introduction}
\begin{figure}
  \centering
  \includegraphics[width=0.5\textwidth,keepaspectratio,height=4cm]{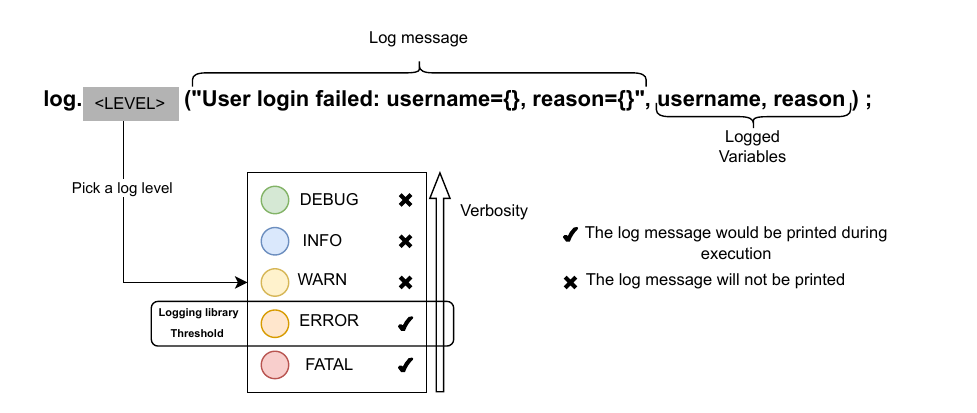}
  \caption{Logging statement template}
    \vspace{-6mm}
  \label{fig:log_stmnt}
\end{figure}

Developers insert logging statements into the source code of software systems to capture essential runtime information, facilitating maintenance, debugging, and system observability~\cite{Li:2020,Shang:2015,Yuan:2012}. A typical logging statement includes a function call specifying the verbosity level (e.g., DEBUG), a textual message, and variables providing contextual information, as shown in Figure~\ref{fig:log_stmnt}. Along with the threshold log level defined by the logging framework, log levels inserted into logging statements --and that are ordered by verbosity from DEBUG (most verbose) to FATAL (least verbose) for most Java libraries-- control the volume of logged data, directly impacting software performance and maintainability. Logging information serves multiple stakeholders, including developers diagnosing issues, DevOps teams monitoring systems, and release managers assessing deployments.

%However, determining the appropriate log level for each logging statement remains a challenging activity for developers~\cite{Li:2017,Pecchia:2015}, and with significant consequences. In fact, insufficient logging reduces observability and harms debugging, while excessive logging leads to performance overhead and overwhelming information

However, determining the appropriate log level for each logging statement remains a challenging activity for developers~\cite{Li:2017,Pecchia:2015}, especially since developers typically make initial poor log level choices~\cite{Li:2020,Oliner:2012}. Suboptimal log level decisions have significant practical consequences as a too low log level drastically reduces software observability, complicates debugging, and prolongs failure diagnosis~\cite{Yuan:2012}, while a too excessive log level introduces performance overhead, inflates storage, and overwhelms developers with redundant information~\cite{Yuan:2014}. To ease the burden on developers, machine learning-based log level predictors (LLPs)~\cite{Anu:2019,Li:2017,Kim:2019,LiZ:2021a} were proposed, including some that leverage large language models (LLMs), and which we refer to in this paper as LLM-LLPs~\cite{Heng:2024,UniLog:2024}. 

While LLMs (e.g., CodeLLama) demonstrate promising results for automating log level selection (AUC ranging from 0.64 to 0.8), the way they are leveraged (e.g., In-Context learning) overlooks the internal structure of software systems. In fact, current LLM-LLPs treat projects as uniform entities (i.e., all components follow the same logging strategies) when retrieving in-context examples for prediction, while in reality, distinct components or subsystems within the same project may exhibit significantly different logging practices, influenced by factors such as developer ownership and code functionality. Ignoring these internal distinctions prevents LLM-LLPs from effectively leveraging the most contextually relevant ICL logging examples, therefore limiting their predictive accuracy and practical utility.

%In a prior work~\cite{Ouatiti:2023}, we investigate the impact of leveraging information from projects' components (as defined by the project's documentation) on LLP performance, demonstrating that logging practices indeed vary across components (e.g., one component using more verbose log levels than another). 

%In a prior work~\cite{Ouatiti:2023}, we observed that logging practices, such as the choice of log verbosity levels, vary significantly across the components (as defined by project's documentation) of the same software system. Yet, many open-source projects lack comprehensive documentation to identify such components. Even if they exist, these static, documentation-based definitions of components (i.e., based on the project structure or docs) might fail to completely capture the diverse functional intents, and the logging styles within software systems. 

For example, the SIG‑Auth team~\footnote{https://github.com/kubernetes/community/tree/master/sig-auth} in the k8s contributing community maintains files across multiple parts of the system (e.g., apiserver vs. kubelet), these files might have similar logging practices, as they are written by the same developers. Similarly, the Hadoop common component contains a retry-handler (\textit{RetryInvocationHandler.java}), which wraps RPC calls in a retry loop, logs each failed attempt at ``DEBUG'', and escalates to ``ERROR'' after the last retry. Meanwhile, the Yarn component contains (\textit{RequestHedgingRMFailoverProxyProvider.java}), a class that performs the same retry duty for Resource-Manager calls, logs attempts at ``DEBUG'', and issues a final ``WARN'' once fail-over succeeds. Such examples demonstrate that logging conventions can be strongly influenced by shared developer ownership and functional similarities. Consequently, retrieving examples without considering these internal distinctions may reduce the contextual relevance of retrieved examples, limiting the predictive accuracy of LLM-LLPs.

% While both files implement the same intent (i.e., automatic retry with escalation), the documentation-based component definition treats them as different components (Common vs. Yarn).

In this paper, we propose a novel framework that enhances the performance of LLM-based LLPs by integrating two complementary clustering approaches: (1) semantic clustering, which groups source code files based on the similarity of their functionality, and (2) ownership clustering, which groups files by shared developer contributions. Finally, we combine these two signals (i.e., semantic and ownership) using a multiplex clustering approach~\footnote{https://leidenalg.readthedocs.io/en/stable/reference.html} to power OmniLLP, our ICL retrieval framework that supplies LLMs used for log level prediction with semantic and ownership-aligned in-context-learning (ICL) logging examples. We hypothesize that these clustering approaches are effective due to the very nature of software engineering practices where files performing similar tasks share semantic patterns, and files maintained by the same developers reflect consistent coding and logging conventions. These clusterings are intended to be updated periodically (e.g., monthly or quarterly) and depending on the activity rate in the software project. We summarize our contribution in the following RQs:

%Additionally, we combine these two signals using a multiplex clustering approach~\footnote{https://leidenalg.readthedocs.io/en/stable/reference.html} to create OmniLLP, a retrieval-augmented generation (RAG) framework that supplies our LLM-LLP with intent and ownership-aligned in-context-learning (ICL) examples for log level prediction. Multiplex clustering is particularly suited to our problem because it allows the simultaneous integration of semantic intent and developer ownership information which are two inherently intertwined aspects of logging practices. For instance, files authored by the same group of developers typically share consistent logging conventions (i.e., ownership clustering), yet these files might span diverse functionalities, potentially reducing semantic cohesion. Conversely, files grouped solely by functionality (i.e., semantic clustering) may neglect the aspect of shared developer logging conventions. By leveraging multiplex clustering, we combine these complementary perspectives, ensuring that the retrieved ICL examples capture both the semantic context and the developer-specific logging conventions. 
 
%We empirically evaluate our clustering approaches on four widely-used open-source software projects: Hadoop, HBase, Cassandra, and Elasticsearch to answer the following research questions:

\textbf{RQ1. \RQI}

%Leveraging ICL examples from our file ownership clusters statistically significantly enhances the precision of LLMs used for log level prediction by a median of 2\% to 7\% (Wilcoxon test, $\alpha =0.01$) compared to the same LLMs when leveraging ICL examples from the whole project. Furthermore, the ownership clustering powering this retrieval mode provides a file coverage (i.e., \% of files that were assigned to a specific ownership-based cluster instead of the noise cluster) ranging between 67\% and 99.5\% across our evaluated projects.

Leveraging ICL examples retrieved from our file ownership clusters statistically significantly enhances the precision of LLM-based log level prediction by a median of 2\% to 7\% (Wilcoxon test, $\alpha=0.01$), compared to using randomly selected ICL examples from the entire project. Additionally, the ownership clustering approach powering this retrieval achieves high file coverage, assigning between 67\% and 99.5\% of the files across our evaluated projects to specific ownership-based clusters rather than the noise cluster.

%the statistically significant precision improvement brought by the ownership-based retrieval compared to the documentation-based retrieval is between 2\% and 3\% (Wilcoxon test, $\alpha =0.01$).

\textbf{RQ2. \RQII}
% Leveraging ICL examples from our code semantic (i.e., code functional intent) clusters statistically significantly enhances the AUC of LLMs used for log level prediction by a median of 4\% to 8\% (Wilcoxon test, $\alpha =0.01$) compared to the same LLMs leveraging ICL examples from the whole project. %Furthermore, the code semantic based retrieval statistically significantly outperforms the retrieval based on documentation-based components (if any) by a median of 5\% to 6\% (Wilcoxon test, $\alpha =0.01$) in terms of AUC. 
% That said, we observe that the file coverage (i.e., \% of files that were assigned to a specific semantic-based cluster instead of the noise cluster) of our semantic-based clustering is between 51\% and 63\% across our studied projects. 

Leveraging ICL examples retrieved from our semantic (i.e., code functional intent) clusters statistically significantly enhances the AUC of LLM-based log level prediction by a median of 4\% to 8\% (Wilcoxon test, $\alpha =0.01$), compared to using random ICL examples from the entire project. That said, we observe that semantic clustering achieves a moderate file coverage, assigning between 51\% and 63\% of files across our studied projects to specific semantic-based clusters rather than the noise cluster.

While the ownership-based clustering offers a near-perfect coverage of files (90\% average coverage) in the studied projects, its performance is inferior to the semantic-based clustering for which coverage is lower (57\% median coverage). In the following RQ we investigate the integration of ownership and semantic signals into a single multiplex clustering, with the aim of creating a retrieval system providing both higher coverage and enhanced performance.

\textbf{RQ3. \RQIII}

Our OmniLLP approach, leveraging the results from multiplex clustering of semantic and ownership signals, achieves superior predictive accuracy, reaching a median AUC between 0.88 and 0.96 across evaluated projects, which represents a median (across projects) increase of 9\% compared to the random retrieval baseline. Furthermore, OmniLLP leverages ownership information to achieve the same near-perfect coverage as the ownership-based approach, therefore guaranteeing significantly less fallbacks to the random retrieval from the whole project data. 

Our findings highlight the importance of considering in-depth software engineering knowledge such as the combination of both semantic intent and developer ownership when automating log level prediction. By providing LLM-LLPs with more coherent context, OmniLLP not only improves predictive performance but also aligns better with the real-world logging practices observed in complex modern software systems.

The remainder of the paper is structured as follows. Section 2 reviews related work. Section 3 details our methodology. Section 4 presents our empirical results. Section 5 discusses threats to validity, and Section 6 concludes the paper.

\section{Background \& Related work}
This paper targets the enhancement of machine learning models used for automating 
log level prediction within multi-component systems. We discuss the following research directions as they are the closest to our work:

\subsection{Log level prediction}
One direction in software logging research aims to automate the log level choice~\cite{Anu:2019,Heng:2024,Kim:2019,Li:2017,LiZ:2021a}, as developers and empirical studies reported challenges with the log level choice activity. For instance, Li et al.~\cite{Li:2017} introduced an approach leveraging an ordinal logistic regression model, specifically designed to predict log levels using a set of static, change-related and historical code features (e.g., log message length, logging statement churn, number of revisions in history). Anu et al.~\cite{Anu:2019} introduced VerbosityLevelDirector, an automated approach based on random-forest designed to help developers assign appropriate log levels based on logging intention embedded within the code context. Kim et al.~\cite{Kim:2019} introduced a new perspective for recommending appropriate log levels based on the semantic and syntactic context of logging statements. Their approach builds a semantic vector that represents the similarity between log levels and the terms associated with such levels, and a syntactic vector capturing structural context surrounding the logging statement in the code. %Using these combined semantic and syntactic vectors, Kim et al.~\cite{Kim:2019} compared four different machine learning classifiers: K-nearest neighbors (KNN), Random Forest, Support Vector Machine (SVM), and Decision Trees for log level prediction. 

More recently, Li et al.~\cite{LiZ:2021a} introduced a deep learning take on log level prediction. Their approach leveraged a neural network architecture that encoded syntactic and semantic contexts surrounding logging statements as well as the contents of the logging statements, to predict an ordinal output corresponding to the different log levels. Heng et al.~\cite{Heng:2024} explored log level prediction using an LLM-based approach, employing various LLMs with different sizes (e.g., CodeLlama-7B). Given a source-code snippet (where the logging statement is masked) as a context, and the contents of the log message, the LLM predicts the appropriate log level by generating an appropriate textual response. 

That said, there is a limited understanding of how the software architecture (e.g., multi-components) influences logging practices, as prior studies typically consider projects as a single coherent entity. In this paper, we investigate log level prediction approaches that take into consideration the multi-component nature of modern software systems. 

\subsection{In-context learning for software engineering tasks}
Recent advancements in large language models (LLMs) have highlighted their adaptability and effectiveness across diverse software engineering tasks~\cite{Jelodar:2025,UniLog:2024,Zhang:2024}. A significant contributor to the flexibility of LLMs is their capability to leverage transfer learning which allows them to apply pretrained knowledge to various downstream tasks. Nonetheless, the generalized knowledge encoded in LLMs often lacks software engineering domain knowledge.

To mitigate this limitation and enhance performance, prior works leveraged In-context learning (ICL). This approach involves prompting LLMs with specific context and examples directly within the input provided to the model, without modifying the model's parameters~\cite{Brown:2020,Wang:2020}. This approach enables models to dynamically adapt the new task output based on given task-specific demonstrations/examples. However, ICL has a major constraint related to the limited context window, which restricts the number of examples that can be provided during inference. This constraint can potentially diminish ICL's effectiveness.

Prior studies have leveraged ICL for logging automation. For instance, Xu et al.~\cite{UniLog:2024} introduced UniLog, an LLM-based solution that employs ICL to automatically generate log messages, recommend appropriate logging statement placements, and predict their verbosity levels. Heng et al.~\cite{Heng:2024} investigated log level prediction using various LLMs with different sizes and model architectures, and showed that even randomly picked ICL examples can lift the performance of LLM-LLPs.

Our work is in line with this research direction as we evaluate ICL approaches for log level prediction. Yet, instead of random ICL retrieval, our study investigates how the selection of ICL examples can be improved when taking into consideration the ownership and semantic relationships present within modern software systems. %Our work directly addresses this gap by proposing a clustering-guided approach to retrieve high quality relevant examples for effective ICL for log level prediction.

\subsection{Local vs. Global ML modeling}
Local machine learning approaches partition software data into subsets exhibiting similar characteristics to train specialized / localized models, instead of considering the entire dataset (i.e., global models). Such subsets can be identified from explicit project documentation (e.g., grouping files documented as the \textit{networking} or \textit{database} modules), or inferred implicitly through clustering techniques (e.g., grouping files by maintenance activity). Prior research has demonstrated the effectiveness of local modeling in several software engineering tasks, including defect prediction, effort estimation, and log level prediction. For instance, Bettenberg et al.\cite{Battenburg:2012} proposed clustering project data to train local defect prediction models, observing improved predictive performance compared to global models that leverage data from the entire project. Menzies et al.\cite{Menzies:2013} proposed local models trained on project-level similarities (i.e., groups of similar project) to improve performance of effort estimation models. Additionally, Ouatiti et al.~\cite{Ouatiti:2023} recommended leveraging local models trained on individual components (explicitly defined by the project's documentation) rather than leveraging data from the entire project for log level prediction.

%While our motivation to leverage localized insights from specific groups within/across a project/ecosystem aligns with these prior studies, our approach introduces a novel solution at inference time rather than relying on training or fine-tuning models. Specifically, our contribution is an informed retrieval framework designed to enhance ICL by selecting contextually relevant examples based on semantic and developer ownership clustering, thereby enabling more precise log level predictions.

While our motivation—to leverage localized insights from specific groups within/across a project/ecosystem aligns with prior studies, our approach differs significantly. Instead of training or fine-tuning specialized models for each identified group, we introduce an informed retrieval framework that enhances ICL by dynamically selecting contextually relevant examples based on semantic and developer ownership signals. By doing so, we enable existing LLMs to achieve more accurate log level predictions without requiring additional model training.

\section{Methodology}

% In this section, we present OmniLLP, our retrieval framework designed to enhance log level prediction for LLM. We first describe the proposed approach and then separately detail our empirical evaluation setup.

% \subsection{OmniLLP framework}
% \label{sec:omni}
In this section, we introduce OmniLLP, as an in-context learning (ICL) retrieval framework that was designed to enhance log level prediction (LLP) performance by automatically identifying and providing relevant logging examples. OmniLLP is intended to seamlessly integrate with LLMs to assist developers selecting the appropriate log level for newly added or modified logging statements within source code.

\begin{figure}[t]
  \centering
  \includegraphics[width=\textwidth,keepaspectratio,height=5.5cm]{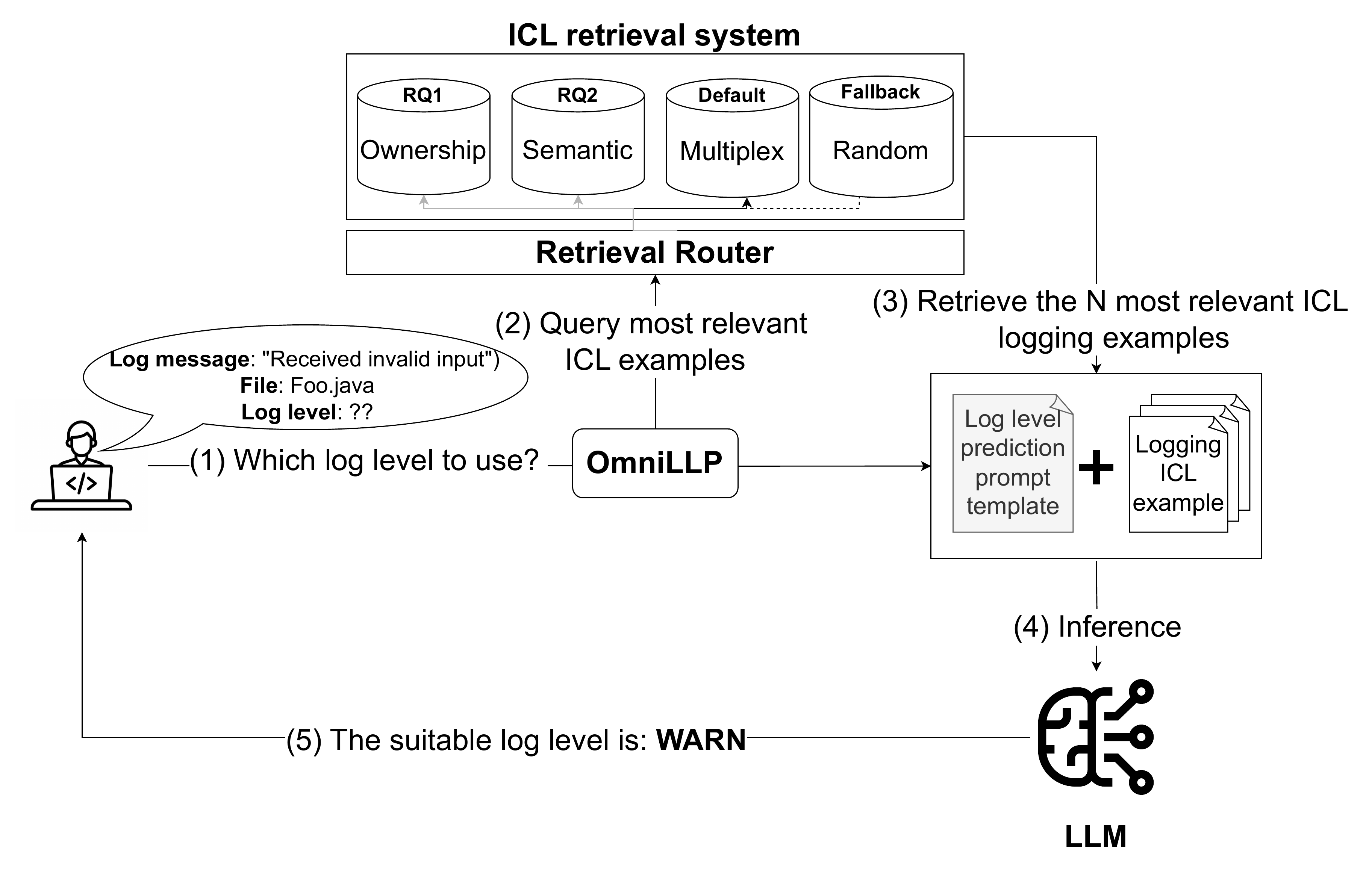}
  \caption{Overview of our OmniLLP framework for log level prediction using ICL retrieval.}
    %\vspace{-6mm}
  \label{fig:methodology_overview}
\end{figure}

To use OmniLLP, a developer simply submits a request containing the file name and the specific logging message for which they require a log level prediction, as shown in Figure~\ref{fig:methodology_overview}. OmniLLP then automatically parses the submitted file, extracting relevant contextual information (i.e., surrounding source code). Using this information, the framework identifies and retrieves the most contextually relevant ICL logging examples, according to the context clustering approach in use (e.g., semantic retrieval). These examples are then combined with the target logging statement and passed to a Large Language Model (LLM), which outputs the predicted log verbosity level.

OmniLLP's retrieval mechanism relies on pre-computed clusters obtained using two complementary clustering techniques: (1) semantic clustering and (2) ownership clustering. We discuss these approaches in detail in the following subsections.

\subsection{Semantic clustering}
\label{sec:sem}
We leverage semantic clustering to group files based on their functional similarity, aiming to identify clusters whose files implement closely related tasks and thus share similar logging behaviors.

\begin{figure}[t]
\centering
\includegraphics[width=0.5\textwidth,keepaspectratio,height=6cm]{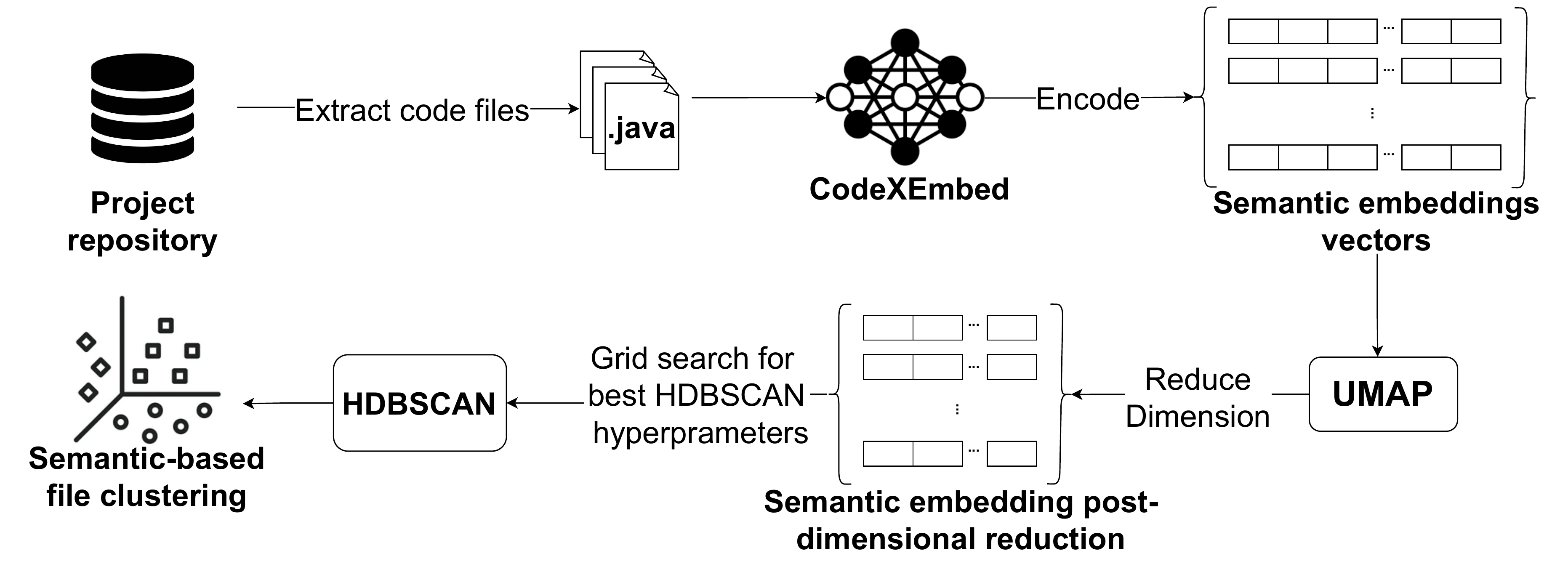}
\caption{Overview of the semantic-based clustering}
%\vspace{-6mm}
\label{fig:semantic_clustering}
\end{figure}

To build our semantic-based clusters we follow the approach summarized in Figure~\ref{fig:semantic_clustering}. First, we create a transformer-based embedding for the files of each project. Such embeddings are dense vector representations obtained by encoding textual inputs (e.g., source code) using transformer models, including encoder-only architectures, such as CodeBERT~\cite{Feng:2020} (encoder-only) and CodeT5~\cite{wang:2021} (encoder-decoder). These models are extensively pre-trained on massive amounts of text and source code data, enabling them to capture semantic patterns within their inputs. Encoder-only models are particularly effective for embedding tasks as their training encourages the model to learn representations that take into account both syntax and semantics~\cite{Chochlov:2022,Feng:2020,Liu:2024}. %These embeddings are typically derived from the final hidden layer outputs of the pre-trained model, and result in vectors where semantically similar inputs are mapped closely in the embedding space.

For the purpose of our study, we chose the encoder-only transformer CodeXEmbed~\cite{Liu:2024} because it provides state-of-the-art performance on code-related embedding tasks, supports input sequences of up to 32K tokens --thus effectively capturing the full context of most source files in our studied projects-- and achieves these capabilities despite its modest size (2.6B parameters), making it computationally efficient and practical for our experiments.

After the encoding, each source file is represented as a 2304-dimensional embedding vector (default dimension for CodeXEmbed embeddings) capturing the file’s semantic information. For exceptionally large files (less than 5\% across projects) exceeding the 32K token window, we partition them at top-level method boundaries, individually encode each method, and then aggregate the resulting vectors via mean-pooling to produce a single representative embedding for the entire file~\cite{Günther:2024}. 

As direct clustering on high-dimensional embeddings is computationally expensive and noisy, we use the UMAP (Uniform Manifold Approximation and Projection)~\cite{Mcinnes:2020} algorithm to project these embeddings into a more manageable 50-dimensional space, similar to prior work~\cite{Baligodugula:2025}. We chose UMAP as the dimensionality reduction approach due to its effectiveness in preserving the relative proximity of semantically related files while significantly reducing computational complexity. 

Next, we leveraged the density-based clustering algorithm HDBSCAN~\cite{Malzer:2020}, which can effectively identify clusters of varying densities and label outliers as noise. To ensure robust and high-quality clusters, we conducted a grid search over a broad hyperparameter space of:
\begin{itemize}
    \item Minimum cluster size (mcs): For every project with N files we tried eight values that cover tiny to large clusters. Specifically, we investigated: N/300 files, N/150 files and a fixed mid-range set {25, 40, 50, 75, 100} in addition to N/10 files (large setting).
    \item Minimum samples (ms): For every mcs value we tried two separate sizes: (1) at $0.5 \times mcs$ and (2) at $0.25 \times mcs$.
\end{itemize}
For every (mcs, ms) pair we computed Silhouette and Davies–Bouldin scores, then selected the setting that maximized Silhouette while simultaneously minimizing Davies–Bouldin, as those are the settings that yield clusters with the best balance of internal cohesion and clear separation from neighboring clusters. Additionally, we validate the stability of the obtained clusters through bootstrap resampling (30 iterations) and computing the Adjusted Rand Index (ARI) across bootstrap iterations.

%Note that our router leverages FAISS~\footnote{\url{https://github.com/facebookresearch/faiss}} to retrieve the five most similar logging examples from within the same semantic-based cluster to serve as in-context examples for log level prediction.

\subsection{Ownership clustering}
\label{sec:own}
We perform ownership clustering to group files according to their shared developer contributions, based on the intuition that files maintained by the same authors tend to exhibit similar logging conventions.

\begin{figure}[t]
  \centering
  \includegraphics[width=0.5\textwidth,keepaspectratio,height=6cm]{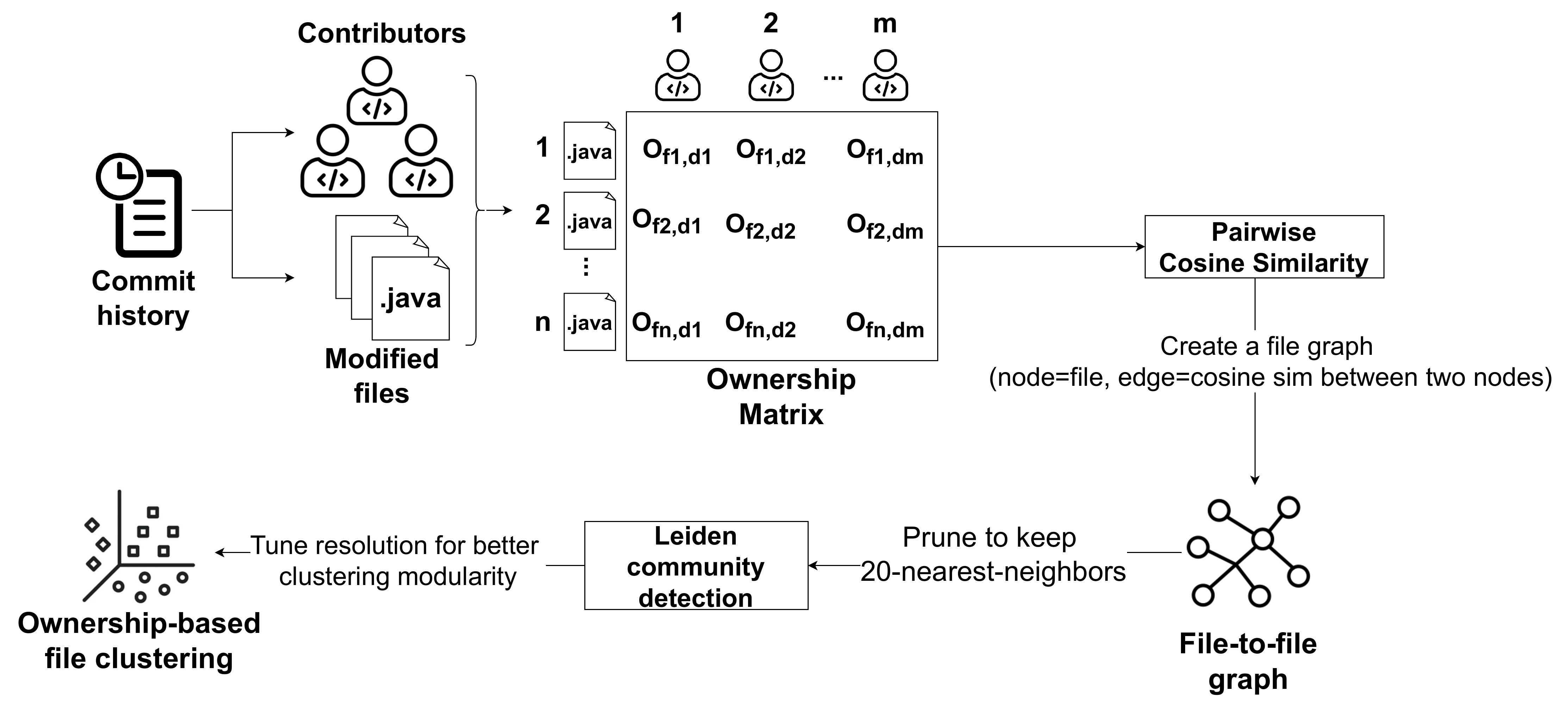}
  \caption{Overview of the ownership-based clustering}
    %\vspace{-6mm}
  \label{fig:ownership_clustering}
\end{figure}

To construct our ownership-based clusters, we mine the Git history of our studied projects and generate an author-file \textit{ownership matrix}, as shown in Figure~\ref{fig:ownership_clustering}. Rows in this matrix represent source files, and columns represent authors. Each matrix cell captures the frequency of file modifications made by an author, weighted by an exponential decay factor to reduce the influence of older changes. 

Treating each row as a high-dimensional ownership vector for its corresponding file, we calculate the cosine similarity for each pair of files. These files along with their simlarity scores, form a weighted file-to-file graph that represents the shared maintenance signal (i.e., connected files are frequently changed by similar author groups). We retain only the 20 strongest cosine-similarity edges per file (i.e., its 20 nearest neighbors), discarding weaker incidental connections (e.g., due to renames or formatting). 

We subsequently apply the weighted Leiden community detection algorithm~\cite{Traag:2019} to this pruned graph. This algorithm is designed for graph-based clustering problems. Specifically, it iteratively improves node partitions through two key phases, providing higher-quality clusters and addressing limitations found in the earlier algorithm Louvain~\cite{Blondel:2008}. In the \textit{first phase} (i.e., local moving phase), each node in the graph is assigned to an initial cluster, after which nodes are iteratively moved to neighboring clusters if such moves improve the modularity score of the partition. This local movement continues until no further improvement is possible, thereby converging to a locally optimal partition. In the \textit{second phase} (refinement phase), the algorithm aggregates clusters found in the first phase into super-nodes, hence, forming a smaller graph where each cluster is represented as a single node. Edges between these super-nodes represent the aggregated connectivity between the original clusters. The algorithm then recursively applies the local moving phase (i.e., first phase) to this condensed graph, further refining and optimizing community partitions. 

These two phases alternate repeatedly until modularity can no longer be improved, resulting in stable and cohesive clustering. The Leiden algorithm's key strength lies in its ability to prevent disconnected clusters, which was a known limitation of the Louvain algorithm, all while providing better scalability and improved modularity scores. As the Leiden algorithm includes a tunable resolution parameter, we experiment with different values to ensure high modularity (greater than 0.7).

%We assess the robustness of the obtained clusters by repeating the clustering procedure over ten non-overlapping temporal windows (2 months-long) spanning the project's history, reporting the Adjusted Rand Index (ARI) to measure stability across these windows.

%Once ownership clusters are determined, we serve our LLP with the closest ICL examples exclusively from the same ownership community as the target file containing the logging statement to be predicted. If no appropriate ownership cluster is identified, we revert to randomly selected ICL examples as a fallback. We evaluate the effectiveness of this approach against two baselines: (a) a single global predictor using examples drawn from the entire project, and (b) local predictors leveraging documentation-based components rather than ownership-based clusters.

\subsection{Multiplex clustering}
\label{sec:mlpx}
\begin{figure}[t]
  \centering
  \includegraphics[width=0.5\textwidth,keepaspectratio,height=3cm]{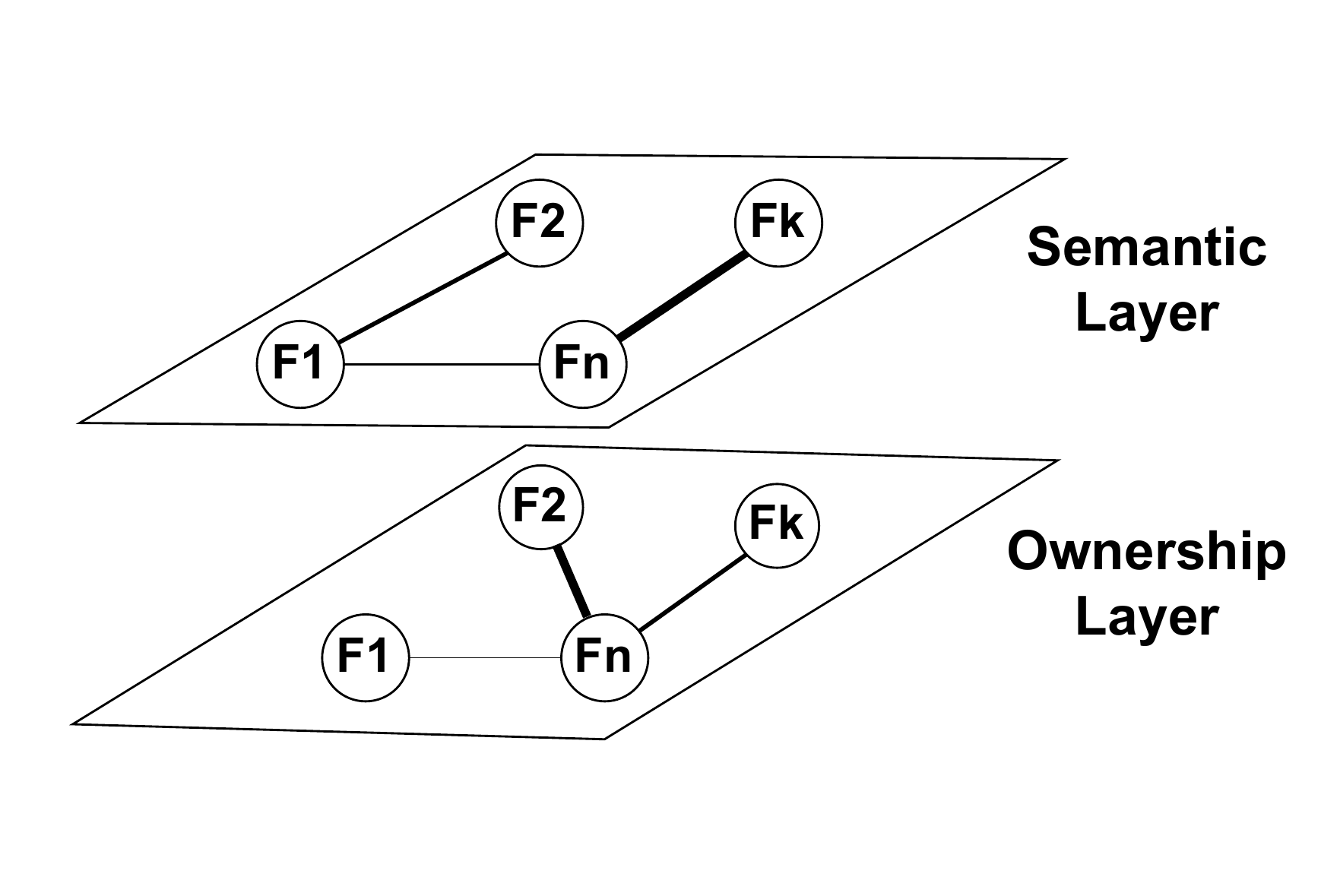}
  \caption{Overview of multiplex graphs. Nodes are Java files and edges are labeled with cosine similarity scores between the vectors representing the connected files.}
    %\vspace{-6mm}
  \label{fig:multiplex_graph}
\end{figure}

As we believe that both semantic similarity and developer ownership signals are crucial for accurate log level decisions, we combine these two complementary clustering strategies into a unified multiplex graph, as shown in Figure~\ref{fig:multiplex_graph}. Specifically, we constructs a two-layer multiplex graph, where each layer independently captures distinct insights. While the first layer encodes \textit{semantic similarity} between files based on code embeddings, the other represents \textit{file ownership} based on developer contributions.

To construct the \textit{semantic similarity} layer, we leverage the transformer based embeddings calculated in Section~\ref{sec:sem} and connect each file to its 20-nearest neighbors using cosine similarity. Edge weights within this semantic layer are subsequently rescaled to the [0,1] interval, normalizing the strength of semantic relationships among files. In parallel, the \textit{developer ownership} layer leverages the exponentially decayed author vectors from Section~\ref{sec:own} to represent files according to their developer contribution patterns. Similar to the semantic layer, we connect each file to its 20-nearest neighbors based on cosine similarity between ownership vectors, then rescale edge weights to the [0,1] range. This ensures a balanced and directly comparable weighting scheme across both semantic and ownership layers.

Both layers have the same set of nodes (each node corresponding to a Java source file) but maintain distinct edge sets, preserving the unique insights offered by each type of signal. To identify cohesive clusters across these complementary layers, we apply the multiplex variant of the Leiden community detection algorithm~\cite{Traag:2019}. This multiplex algorithm simultaneously optimizes modularity across both semantic and ownership layers, identifying clusters where files share not only similar functionality but also similar maintaining developers.

At inference time, for a given logging statement, OmniLLP identifies the multiplex cluster associated with the source file containing the logging statement, then it retrieves the top-5 scoring examples within this cluster, using a combined similarity score defined as: 
\[
s(c) = 0.7 \cdot \cos_{\text{sem}}(f^*, c) + 0.3 \cdot \cos_{\text{own}}(f^*, c)
\]
where: 
\begin{itemize}
\item $s(c)$: is the combined similarity score for candidate example $c$. 
\item $f^*$ : is the file containing the logging level to be predicted. 
\item $cos_{sem}(f^{*},c)$: is the semantic cosine similarity between the embeddings of $f^*$ and candidate $c$. 
\item 
$cos_{own}(f^{*},c)$ is the ownership cosine similarity between the embeddings of $f^*$ and candidate $c$. 

\end{itemize}

For the purposes of this study, we assign a higher weight to the semantic similarity score due to its superior predictive improvements compared to ownership similarity, as demonstrated in Section~\ref{sec:results}. However, both weights are configurable parameters in OmniLLP, enabling end-users to adjust their relative importance based on their own requirements.

\subsection{Operational workflow of OmniLLP}
OmniLLP can operate in three distinct modes: (1) semantic-only, (2) ownership-only, and (3) multiplex, which integrates both signals simultaneously (default mode). Additionally, if OmniLLP cannot identify a suitable cluster or if the cluster does not contain enough logging examples, based on the selected mode and the input logging statement, our framework automatically falls back to randomly retrieving examples from the entire project dataset, thus ensuring baseline predictive performance at all times.

All our presented clustering approaches are computationally efficient, ensuring practical applicability for cluster refreshing if/when needed. Specifically, ownership-based clustering completes in a few minutes, even for our largest studied project (Elasticsearch). Semantic-based clustering takes slightly longer due to the initial step of computing embeddings using a transformer-based tokenizer. However, it remains practical, taking approximately 20 minutes for Elasticsearch (the largest project studied) on standard hardware (16-core CPU, 64 GB RAM, 24 GB GPU memory). Finally, once the artifacts from the prior two clusterings (i.e., ownership matrix and semantic embeddings) are created, multiplex clustering takes few minutes for all of our projects. 

For the purposes of this study, we chose CodeLLama-7B due to its optimal balance between predictive performance (best performance based on prior study~\cite{Heng:2024}) and manageable size, facilitating practical deployment even on edge devices (e.g., a developer's laptop), unlike larger models that require more compute~\cite{Kaplan:2020}. 

% Finally, to predict a log level for a new logging statement, we adopt a prompting approach similar to prior work~\cite{Heng:2024} where we extract the surrounding source code from the source file containing the logging statement. We ensure the removal of logging guards (e.g., \textit{if (logger.isDebugEnabled())}) to prevent data leakage. Next, we mask the logging statement to simulate a real-world prediction scenario. We then construct a prompt, combining the logging statement surrounding source code as context with the extracted log message. The prompt instructs the LLM to predict the appropriate log level from a predefined set of verbosity options (e.g., debug, info, warn, error). Finally, the LLM generates the predicted log level as output. 

% \begin{figure*}
% \centering
% \includegraphics[width=\textwidth,keepaspectratio,height=4cm]{figures/llm-llp.pdf}
% \caption{Construction of the prompt for log level prediction~\cite{Heng:2024}}
% \label{fig:llm_prediction}
% \end{figure*}

\section{Empirical evaluation setup}
We empirically evaluate the effectiveness of OmniLLP (with its different operation modes) across four widely used open-source Java projects: Hadoop, HBase, Elasticsearch, and Cassandra. These projects were selected due to their varied size, distinct logging practices, and differing clarity in documentation-based component definitions. Specifically, Hadoop and HBase feature explicit documentation-defined components, allowing comparison against a documentation-based retrieval baseline. Conversely, Elasticsearch and Cassandra lack clearly defined component boundaries, presenting a scenario ideal for assessing OmniLLP's robustness and generalizability.

\subsection{Data collection}

We collected the datasets used in this paper by extracting source code files, logging statements, and associated metadata directly from each project's GitHub repository. Specifically, we cloned each repository and checked out the latest stable release versions (e.g., Cassandra-4.1.9) to ensure reproducibility. To extract logging statements systematically, we parsed Java source files using the \textit{javalang}\footnote{https://pypi.org/project/javalang/} library, identifying log invocations and capturing relevant context such as the log message, verbosity level, method boundaries, and surrounding code snippets. Metadata such as file paths, line numbers, and documentation-based components (if any) were also recorded. In cases where a project lacked explicit documentation-based component definitions (e.g., Cassandra), we labeled the component information as \textit{unknown}. Finally, the extracted logging data was stored in a MongoDB database to facilitate efficient querying in our following analysis.
Table~\ref{tab:project_details} summarizes the characteristics of each studied project.

Table~\ref{tab:project_details} summarizes key characteristics of each of our studied projects.
\begin{table}[]
\resizebox{0.45\textwidth}{!}{\begin{tabular}{|l|l|l|l|l|}
\hline
\textbf{Project} & \textbf{Versions} & \textbf{\# SLOC} & \textbf{\# Contributors} & \textbf{\# Logging statements} \\ \hline
Hadoop           & 3.3.6             & 1.8M             & 635                      & 21,910                          \\ \hline
HBase            & 2.5.9             & 841K             & 483                      & 10,449                          \\ \hline
Elasticsearch    & 8.9.2             & 2.5M             & 2,042                     & 25,187                          \\ \hline
Cassandra        & 4.1.9             & 595K             & 464                      & 4,808                           \\ \hline
\end{tabular}}
\caption{Summary of the studied projects}
\label{tab:project_details}
\end{table}

% \subsubsection{Prompting LLMs for log level}
% To predict a log level, we adopt a prompting approach similar to prior work~\cite{Heng:2024}. First, we extract the log message, actual log level, and surrounding source code from the source file containing the logging statement. We ensure the removal of logging guards (e.g., \textit{if (logger.isDebugEnabled())}) to prevent data leakage. Next, we mask the logging statement to simulate a real-world prediction scenario. We then construct a prompt, combining the logging statement surrounding source code as context with the extracted log message. The prompt instructs the LLM to predict the appropriate log level from a predefined set of verbosity options (e.g., debug, info, warn, error). Finally, the LLM generates the predicted log level as output. 

% % \begin{figure*}
% % \centering
% % \includegraphics[width=\textwidth,keepaspectratio,height=4cm]{figures/llm-llp.pdf}
% % \caption{Construction of the prompt for log level prediction~\cite{Heng:2024}}
% % \label{fig:llm_prediction}
% % \end{figure*}

% For the purposes of this study, we chose CodeLLama-7B due to its optimal balance between predictive performance (best performance based on prior study~\cite{Heng:2024}) and manageable size, facilitating practical deployment even on edge devices (e.g., a developer's laptop), unlike larger models that require more compute~\cite{Kaplan:2020}. 

\subsection{Evaluation baselines}
\label{sec:base}
To evaluate the performance of OmniLLP across its different modes, we leverage two baseline retrieval approaches used by prior studies: 
\begin{itemize}
    \item \textbf{Random retrieval}~\cite{Heng:2024}: Logging examples for ICL are randomly selected from the entire dataset of the studied project.
    \item \textbf{Documentation-based retrieval}~\cite{Ouatiti:2023}: For projects explicitly defining components in their documentation (e.g., HDFS vs. Yarn in Hadoop), logging examples for ICL are exclusively retrieved from within the same documented component (e.g., for a file in Yarn, examples are retrieved only from other Yarn files). This approach relies solely on predefined documentation-based boundaries without employing clustering and is only applicable when such explicit documentation is available.
\end{itemize}

Our evaluation of OmniLLP against the two baselines begins by splitting each project's logging data into retrieval and test sets (70\% retrieval, 30\% test). To robustly assess predictive performance, we evaluated each retrieval method using five bootstrap samples drawn from the test dataset. To determine whether observed performance differences were statistically significant, we employed paired Wilcoxon signed-rank tests, as predictions from OmniLLP and baseline models are directly comparable for each individual logging statement. Additionally, we measured the magnitude of these differences using Cohen's d, interpreting the effect sizes following standard thresholds: small ($0.2 \le d <0.5$), medium ($0.5 \le d < 0.8$), and large ($d \ge 0.8$)~\cite{Cohen:1992}.

\subsection{Evaluation metrics}
As our approach integrates two modeling tasks (i.e., clustering and log level prediction), we evaluate the effectiveness of our method using two distinct sets of robust metrics: clustering quality metrics and log level prediction performance metrics.

\noindent\textbf{Clustering quality metrics}

% \noindent We assess clustering quality using three metrics: Adjusted Rand Index (ARI), Silhouette score, and Leiden modularity.
\begin{itemize}
    \item Adjusted Rand Index~\footnote{\url{https://scikit-learn.org/stable/modules/generated/sklearn.metrics.adjusted_rand_score.html}} (ARI): quantifies the agreement or similarity between two clustering solutions. ARI ranges between -1 and 1, where values close to 1 indicate strong agreement, values close to 0 indicate random clustering, and negative values indicate worse-than-random clustering~\cite{Steinley:2004}.
    \item Silhouette score~\footnote{\url{https://scikit-learn.org/stable/modules/generated/sklearn.metrics.silhouette_score.html}}: measures how tightly grouped (i.e., cohesive) points within clusters are compared to points in other clusters. It ranges from -1 to 1, with values above 0.5 indicate strong, cohesive, and well-separated clusters; values between 0.25 and 0.5 reflect moderate clustering structure; and values below 0.25 suggest weak or ambiguous clustering~\cite{Rousseeuw:1987}. 
    \item Davies–Bouldin index~\cite{Davies:1979} (DBI): measures how well clusters are separated by comparing each cluster’s internal dispersion with its nearest-neighbour distance. It ranges from 0 to $\infty$, where lower values are better. Typically, DBI scores below 1 indicate well-separated and compact clusters~\cite{Davies:1979}. 
    \item Density-based clustering validation~\cite{Moulavi:2014} (DBCV): a density-aware tailored for algorithms such as HDBSCAN. It combines cluster compactness (how tightly points are packed around their local density mode) with cluster isolation (how deep the density drop is between neighbouring clusters). The values range from –1 to +1, with positive scores indicating well-formed, well-separated density regions. Meanwhile, values near 0 suggest structures that are indistinguishable from noise and negative scores indicate badly fragmented or overlapping clusters~\cite{Moulavi:2014}.
    \item  Leiden modularity~\cite{Traag:2019}: assesses how a graph can be partitioned into distinct clusters by providing insight into the overall strength and quality of cluster boundaries. High modularity indicates that nodes within clusters have dense internal connections and relatively sparse connections to nodes in other clusters. Values above 0.3 typically indicate a strong and meaningful community structure ~\cite{Fortunato:2007,Traag:2019} .
\end{itemize}

\noindent\textbf{Log level prediction performance metrics}

% \noindent We evaluate the predictive performance of our LLP models using Area Under the Receiver Operating Characteristic Curve (AUC), Accuracy, and Average Ordinal Distance (AOD):
\begin{itemize}
    \item Area Under the ROC Curve (AUC): quantifies the ability of the model to discrimnate between the different possible classes. Similar to prior works~\cite{Heng:2024,Li:2017,LiZ:2021a,Ouatiti:2023}, we leverage the multi-class variant of AUC~\footnote{\url{https://scikit-learn.org/stable/modules/generated/sklearn.metrics.roc_auc_score.html}} to accommodate our multi-class output. 
    %For each logging statement, our LLM outputs a probability distribution over log levels, obtained by constraining the model to predict a single token corresponding to each verbosity level (e.g., "debug", "info"). These token probabilities are normalized via softmax to represent confidence scores for each possible log level. We then use these probability vectors, along with the true labels (actual log levels), to compute the multi-class AUC, reflecting the model’s ability to distinguish between log level categories effectively.
    \item Precision: measures the fraction of logging statements for which the predicted log level exactly matches the true log level.
    \item Average Ordinal Distance (AOD): quantifies how close the predicted log levels are to their true ordinal positions (i.e., the ground truth log level). A higher AOD indicates predictions that are consistently closer to the true log level labels. Similar to prior work~\cite{Heng:2024} we leverage this metric as it provides a more nuanced assessment of ordinal prediction quality than traditional accuracy.
\end{itemize}
\section{Results}
\label{sec:results}
\subsection*{\textbf{RQ1. \RQI}}

\textbf{Motivation:} Project ownership structure (i.e., who authors the code) often shapes logging practices. For example, the k8s SIG‑Auth team\footnote{https://github.com/kubernetes/community/tree/master/sig-auth} maintains files that reside in different components (e.g., \texttt{apiserver} vs. \texttt{kubelet}). Despite being placed in distinct parts of the system,  these files typically share logging conventions because they are written by the same developers. Consequently, randomly retrieving logging examples without considering ownership structure may result in contextually irrelevant examples, thus limiting the predictive accuracy of LLM-LLPs. By clustering files based on ownership (i.e., grouping files modified by similar sets of developers), we aim to uncover ownership-aligned clusters characterized by consistent log level choice conventions. In this RQ, we investigate whether leveraging ownership-based clusters improves the predictive performance of LLM-LLPs.

\noindent\textbf{Approach:} To quantify the impact of developer ownership information on the performance of LLM-LLPs, we first perform ownership-based clustering for each studied project, as described in Section~\ref{sec:own}. Then, we evaluate the performance of OmniLLP operating in \textit{ownership-only} mode against our two baselines (Section~\ref{sec:base}): (a) random retrieval from the entire project's dataset, and, when available, (b) documentation-based retrieval. 

\textbf{Results}:\textbf{ Our ownership-based clustering approach produces cohesive and stable clusters across all studied projects}, as shown in Table~\ref{tab:ownership_clustering}. In fact, our ownership-based clusters achieve consistently high Leiden modularity scores ($Q>0.79$), with Elasticsearch reaching 0.92. Additionally, intra-cluster cosine similarities (ranging from 0.62 to 0.85) significantly exceed inter-cluster similarities (0.02–0.27), reinforcing the internal cohesion of the identified clusters. Furthermore, the stability analysis of our clustering via the Adjusted Rand Index (ARI) across 15 two-month windows indicates a moderate to high cluster consistency (median ARI between 0.29 and 0.59). Although HBase exhibits relatively low stability (ARI of 0.29), its high modularity (0.81) suggests that while ownership may shift frequently, recalculating clusters periodically (e.g., monthly) would effectively sustain cohesive and useful clusters.

\begin{table}[]
\resizebox{0.5\textwidth}{!}{
\begin{tabular}{|l|l|l|l|l|}
\hline
\textbf{Project} & \textbf{Modularity Qc} & \textbf{Median ARI} & \textbf{Inter/Intra Cosine similarity} & \textbf{Coverage} \\ \hline
Hadoop           & 0.798                  & 0.524               & 0.851 / 0.123                          & 99\%              \\ \hline
HBase            & 0.805                  & 0.286               & 0.618 / 0.018                          & 67\%              \\ \hline
Elasticsearch    & 0.850                  & 0.556               & 0.720 / 0.268                          & 99.5\%            \\ \hline
Cassandra        & 0.919                  & 0.588               & 0.651 / 0.093                          & 96\%              \\ \hline
\end{tabular}
}
\caption{Ownership-based clustering quality metrics}
\label{tab:ownership_clustering}
\end{table}

\textbf{Leveraging ownership-based retrieval significantly enhances the predictive performance across all projects}, as shown in Table~\ref{tab:full_ownership}. Specifically, providing the LLM used for log level prediction with five ownership-based ICL examples statistically significantly (Wilcoxon test, $\alpha =0.01$, Cohen's d, $d>0.8$) improves the median AUC of that LLM from a median 0.81 (observed when using random retrieval) to between 0.83 (Hadoop) and 0.90 (Elasticsearch), marking a median improvement of 2.1\%. Moreover, ownership-based retrieval outperforms (Wilcoxon test, $\alpha =0.01$, Cohen's d, $d>0.7$) retrieval based on documentation-defined component in one of the two projects for which explicit components are defined in the project's documentation. The improvements brought by ownership-based retrieval extend to precision and AOD scores, for which we report improvements from 1\% to 8\% (Precision) and from 0.3\% to 3\% (AOD). 

Our reported scores account for fallback instances where log level predictions are made for files not assigned to any ownership clusters. However, the fallback rate remains minimal (below 5\%) across all studied projects except HBase, which experiences a higher fallback rate of 33\%. This higher fallback rate indicates that many logging statements fall into files that either belong to small clusters (discarded as noise due to our minimum cluster size constraint of 10 files) or clusters lacking sufficient examples for complete ownership-only retrieval. Consequently, we observe the lowest performance improvement (0.3\%) among our projects for HBase when using the \textit{ownership-only} retrieval mode.

% \usepackage{multirow}
\begin{table*}[]
\resizebox{\textwidth}{!}{\begin{tabular}{|l|cll|lcc|ccc|ccc|}
\hline
\multicolumn{1}{|c|}{\multirow{2}{*}{\textbf{Project}}} & \multicolumn{3}{c|}{\textbf{Zero-shot}}                                                                      & \multicolumn{3}{c|}{\textbf{\begin{tabular}[c]{@{}c@{}}Global \\ Few-shot (5)\end{tabular}}}       & \multicolumn{3}{c|}{\textbf{\begin{tabular}[c]{@{}c@{}}Documentation component \\ Few-shot (5)\end{tabular}}} & \multicolumn{3}{c|}{\textbf{\begin{tabular}[c]{@{}c@{}}Ownership-based \\ Few-shot (5)\end{tabular}}}                         \\ \cline{2-13} 
\multicolumn{1}{|c|}{}                                  & \multicolumn{1}{l|}{AUC}           & \multicolumn{1}{l|}{Precision}     & AOD                                & \multicolumn{1}{l|}{AUC}           & \multicolumn{1}{l|}{Precision}     & \multicolumn{1}{l|}{AOD} & \multicolumn{1}{l|}{AUC}                     & \multicolumn{1}{l|}{Precision}      & \multicolumn{1}{l|}{AOD} & \multicolumn{1}{l|}{AUC}                    & \multicolumn{1}{l|}{Precision}                       & \multicolumn{1}{l|}{AOD} \\ \hline
Hadoop                                                  & \multicolumn{1}{l|}{0.694 ± 0.002} & \multicolumn{1}{l|}{0.342 ± 0.001} & 0.729 ± 0.002                      & \multicolumn{1}{l|}{0.802 ± 0.004} & \multicolumn{1}{l|}{0.500 ± 0.008} & 0.806 ± 0.002            & \multicolumn{1}{c|}{0.808 ± 0.004}           & \multicolumn{1}{c|}{0.504 ± 0.004}  & 0.806 ± 0.003            & \multicolumn{1}{c|}{\textbf{0.832 ± 0.002}} & \multicolumn{1}{c|}{\textit{\textbf{0.532 ± 0.002}}} & \textbf{0.819 ± 0.001}   \\ \hline
HBase                                                   & \multicolumn{1}{c|}{0.677 ± 0.002} & \multicolumn{1}{c|}{0.332 ± 0.023} & \multicolumn{1}{c|}{0.732 ± 0.011} & \multicolumn{1}{c|}{0.805 ± 0.000} & \multicolumn{1}{c|}{0.502 ± 0.011} & 0.805 ± 0.003            & \multicolumn{1}{c|}{\textbf{0.813 ± 0.005}}  & \multicolumn{1}{c|}{0.483 ± 0.007}  & 0.800 ± 0.001            & \multicolumn{1}{c|}{0.808 ± 0.002}          & \multicolumn{1}{c|}{\textbf{0.510 ± 0.008}}          & \textbf{0.808 ± 0.002}   \\ \hline
Elasticsearch                                           & \multicolumn{1}{c|}{0.732 ± 0.003} & \multicolumn{1}{c|}{0.400 ± 0.002} & \multicolumn{1}{c|}{0.751 ± 0.001} & \multicolumn{1}{c|}{0.844 ± 0.010} & \multicolumn{1}{c|}{0.496 ± 0.003} & 0.814 ± 0.001            & \multicolumn{1}{c|}{-}                       & \multicolumn{1}{c|}{-}              & -                        & \multicolumn{1}{c|}{\textbf{0.904 ± 0.001}} & \multicolumn{1}{c|}{\textbf{0.578 ± 0.006}}          & \textbf{0.845 ± 0.001}   \\ \hline
Cassandra                                               & \multicolumn{1}{c|}{0.675 ± 0.010} & \multicolumn{1}{l|}{0.339 ± 0.010} & 0.723 ± 0.006                      & \multicolumn{1}{l|}{0.825 ± 0.007} & \multicolumn{1}{c|}{0.481 ± 0.017} & 0.810 ± 0.008            & \multicolumn{1}{c|}{-}                       & \multicolumn{1}{c|}{-}              & -                        & \multicolumn{1}{c|}{\textbf{0.837 ± 0.007}} & \multicolumn{1}{c|}{\textbf{0.501 ± 0.019}}          & \textbf{0.824 ± 0.008}   \\ \hline
\end{tabular}}
\caption{Summary of the performance of the \textit{ownership-only} mode of OmniLLP compared to other approaches (metrics are obtained from 5 bootstrap runs)}
\label{tab:full_ownership}
\end{table*}

\begin{tcolorbox}[colback=gray!10, colframe=black, title=\textbf{Summary of RQ1}]
Leveraging file ownership information enhances retrieval quality for LLM-LLPs, as OmniLLP's \textit{ownership-only} mode achieves a statistically significant improvement in AUC performance, ranging between 0.3\% and 6\%, compared to random retrieval.
\end{tcolorbox}

\subsection*{\textbf{RQ2. \RQII}}

\textbf{Motivation}: Code snippets implementing similar functional intents often exhibit similar logging practice. For example, in Hadoop, the Common component includes a retry-handler (RetryInvocationHandler.java) that wraps Remote Procedure Calls (RPC) in a retry loop, logging each failed attempt at a ```DEBUG''' level, and eventually escalating to ```ERROR''' at the final failure. Meanwhile, the Yarn hosts a semantically similar retry-handler (RequestHedgingRMFailoverProxyProvider.java) performing the same retry logic for Resource Manager calls, logging retries at ```DEBUG''' but escalating only to ```WARN''' upon successful failover. Although these files share highly similar functional and logging intents (automatic retries with escalation), random retrieval approaches are likely to overlook this semantic relationship when selecting ICL logging examples for log level prediction. Such missed semantic similarities can prevent log level predictors from accessing the most relevant examples. In this RQ, we investigate whether leveraging semantic-based clustering to explicitly identify these semantic relationships can improve the predictive performance of LLMs used for log level prediction.

\textbf{Approach}: To quantify the impact of code semantic information on the performance of LLM-LLPs, we first perform semantic-based clustering for each studied project as described in Section~\ref{sec:sem}. Then, we evaluate the performance of OmniLLP operating in "\textit{semantic-only}" mode against three baselines: (a) a global predictor retrieving examples randomly from the entire project, when available, (b) a local predictor that retrieves examples based on explicitly defined, documentation-based components, and (c) OmniLLP's "\textit{ownership-only}" mode. 

\textbf{Results}:\textbf{ Our semantic-based clustering yields compact, cohesive, and stable clusters}, as shown in Table~\ref{tab:semantic_clustering}. In fact, our clustering approach consistently identifies between 27 and 31 semantically aligned clusters, across our studied projects. These clusters achieve a strong internal cohesion as the median Silhouette score ranges from 0.53 to 0.6 (more than 0.5 is cohesive) and the DBCV between 0.23 and 0.35 (less than 1 is compact). Furthermore, stability of our clusters is high, as demonstrated by the bootstrapped ARI scores ranging from 0.67 (Elasticsearch) to 0.7 (Cassandra). That said, we observe that HDBSCAN marks 37\% (Cassandra) to 49\% (Hadoop) of the files as noise, potentially leading to fallback scenarios during retrieval (i.e., if the log level to predicted is in a file labeled as noise).

%A closer analysis of files labeled as noise by HDBSCAN reveals that they predominantly represent unique, specialized functionality, which justifies their exclusion from the semantic clusters. For example, in our inspection of 40 randomly selected noise-labeled files per project, we observe files like Hadoop’s \texttt{QuickSort.java}, which implements generic sorting logic, and Elasticsearch’s \texttt{BlobStore.java}, an abstract representation for binary storage systems. Similarl, Cassandra's \texttt{CacheService.java}, provides cache management functionalities, and HBase’s \texttt{ImmutableBytesWritable.java}, is a utility for handling immutable byte arrays. Such files offer utility methods that are cross-concern and might not align clearly with specific execution intents captured by semantic clusters.

\begin{table}[]
\resizebox{0.5\textwidth}{!}{
\begin{tabular}{|l|l|l|l|l|l|l|l|}
\hline
\textbf{Project}       & \multicolumn{1}{c|}{\textbf{Files}} & \multicolumn{1}{c|}{\textbf{Best mcs / ms}} & \multicolumn{1}{c|}{\textbf{\# Clusters}} & \multicolumn{1}{c|}{\textbf{Silh.}} & \multicolumn{1}{c|}{\textbf{DBCV}} & \multicolumn{1}{c|}{\textbf{Noise \%}} & \multicolumn{1}{c|}{\textbf{Bootstr. ARI}} \\ \hline
\textbf{Hadoop}        & 7 395                               & 50 / 25                                     & 31                                        & 0.533                               & 0.230                              & 49 \%                                  & 0.685 ± 0.027                                \\ \hline
\textbf{HBase}         & 2 691                               & 25 / 12                                     & 27                                        & 0.566                               & 0.325                              & 41 \%                                  & 0.69 ± 0.04                                \\ \hline
\textbf{Elasticsearch} & 14 976                              & 100 / 25                                    & 31                                        & 0.553                               & 0.219                              & 45 \%                                  & 0.67 ± 0.05                                \\ \hline
\textbf{Cassandra}     & 3 158                               & 25 / 12                                     & 30                                        & 0.606                               & 0.347                              & 37 \%                                  & 0.70 ± 0.03                                \\ \hline
\end{tabular}}
\caption{Semantic-based clustering quality metrics}
\label{tab:semantic_clustering}
\end{table}

\textbf{Despite the relatively high ratio of files labeled as noise, leveraging semantic-based clusters significantly enhances LLM-LLP prediction compared to documentation-based and global baselines}, as shown in Table~\ref{tab:full_semantic}. Indeed, using semantic-based retrieval to provide ICL logging examples to our LLM-LLP statistically significantly improves the median AUC scores by 4\% to 6\% compared to the documentation based ICL retrieval (if any), and by 2\% (Elasticsearch) to 6\% (Hbase) compared to ownership-based retrieval. The performance improvement brought by semantic-based retrieval extends to both the precision and AOD scores, across all our studied projects. 

The relatively high portion of files labeled as noise (shown in Table~\ref{tab:semantic_clustering}) is an expected outcome of using HDBSCAN~\cite{Malzer:2020,saha:2023}, as this clustering technique explicitly identifies points in sparse regions of the embedding space as noise. These noise-labeled files in our projects often correspond to specialized or unique implementations that lack sufficient semantic similarity to form dense clusters. Our manual inspection of randomly selected noise files across projects consistently reveals such specialized functionality: for instance, Hadoop’s generic sorting utility (\textit{QuickSort.java}), Elasticsearch’s binary storage abstraction (\textit{BlobStore.java}), Cassandra’s caching mechanism (\textit{CacheService.java}), and HBase’s immutable byte-array handler (\textit{ImmutableBytesWritable.java}) each represent distinct, project-specific functionalities. Furthermore, our statistically significantly (Wilcoxon test, $\alpha =0.01$, Cohen's d, $d>0.7$) improved LLP performance (despite the fallback retrieval mechanism triggered by noise files) indicates that the files identified as non-noise form cohesive clusters, that capture the core functional intents within each project. 

% \usepackage{multirow}
\begin{table*}[]
\resizebox{\textwidth}{!}{\begin{tabular}{|l|cll|lcc|ccc|ccc|}
\hline
\multicolumn{1}{|c|}{\multirow{2}{*}{\textbf{Project}}} & \multicolumn{3}{c|}{\textbf{Zero-shot}}                                                                      & \multicolumn{3}{c|}{\textbf{\begin{tabular}[c]{@{}c@{}}Global \\ Few-shot (5)\end{tabular}}}       & \multicolumn{3}{c|}{\textbf{\begin{tabular}[c]{@{}c@{}}Documentation component \\ Few-shot (5)\end{tabular}}} & \multicolumn{3}{c|}{\textbf{\begin{tabular}[c]{@{}c@{}}Semantic-based \\ Few-shot (5)\end{tabular}}}                          \\ \cline{2-13} 
\multicolumn{1}{|c|}{}                                  & \multicolumn{1}{l|}{AUC}           & \multicolumn{1}{l|}{Precision}     & AOD                                & \multicolumn{1}{l|}{AUC}           & \multicolumn{1}{l|}{Precision}     & \multicolumn{1}{l|}{AOD} & \multicolumn{1}{l|}{AUC}               & \multicolumn{1}{l|}{Precision}         & \multicolumn{1}{l|}{AOD}    & \multicolumn{1}{l|}{AUC}                    & \multicolumn{1}{l|}{Precision}                       & \multicolumn{1}{l|}{AOD} \\ \hline
Hadoop                                                  & \multicolumn{1}{l|}{0.694 ± 0.002} & \multicolumn{1}{l|}{0.342 ± 0.001} & 0.729 ± 0.002                      & \multicolumn{1}{l|}{0.802 ± 0.004} & \multicolumn{1}{l|}{0.500 ± 0.008} & 0.806 ± 0.002            & \multicolumn{1}{c|}{0.808 ± 0.004}     & \multicolumn{1}{c|}{0.504 ± 0.004}     & 0.806 ± 0.003               & \multicolumn{1}{c|}{\textbf{0.865 ± 0.002}} & \multicolumn{1}{c|}{\textit{\textbf{0.587 ± 0.002}}} & \textbf{0.837 ± 0.003}   \\ \hline
HBase                                                   & \multicolumn{1}{c|}{0.677 ± 0.002} & \multicolumn{1}{c|}{0.332 ± 0.023} & \multicolumn{1}{c|}{0.732 ± 0.011} & \multicolumn{1}{c|}{0.805 ± 0.000} & \multicolumn{1}{c|}{0.502 ± 0.011} & 0.805 ± 0.003            & \multicolumn{1}{c|}{0.813 ± 0.005}     & \multicolumn{1}{c|}{0.483 ± 0.007}     & 0.800 ± 0.001               & \multicolumn{1}{c|}{\textbf{0.863 ± 0.003}} & \multicolumn{1}{c|}{\textbf{0.568 ± 0.007}}          & \textbf{0.838 ± 0.004}   \\ \hline
Elasticsearch                                           & \multicolumn{1}{c|}{0.732 ± 0.003} & \multicolumn{1}{c|}{0.400 ± 0.002} & \multicolumn{1}{c|}{0.751 ± 0.001} & \multicolumn{1}{c|}{0.844 ± 0.010} & \multicolumn{1}{c|}{0.496 ± 0.003} & 0.814 ± 0.001            & \multicolumn{1}{c|}{-}                 & \multicolumn{1}{c|}{-}                 & -                           & \multicolumn{1}{c|}{\textbf{0.924 ± 0.002}} & \multicolumn{1}{c|}{\textbf{0.638 ± 0.008}}          & \textbf{0.863 ± 0.001}   \\ \hline
Cassandra                                               & \multicolumn{1}{c|}{0.675 ± 0.010} & \multicolumn{1}{l|}{0.339 ± 0.010} & 0.723 ± 0.006                      & \multicolumn{1}{l|}{0.825 ± 0.007} & \multicolumn{1}{c|}{0.481 ± 0.017} & 0.810 ± 0.008            & \multicolumn{1}{c|}{-}                 & \multicolumn{1}{c|}{-}                 & -                           & \multicolumn{1}{c|}{\textbf{0.863 ± 0.006}} & \multicolumn{1}{c|}{\textbf{0.652 ± 0.020}}          & \textbf{0.868 ± 0.005}   \\ \hline
\end{tabular}}
\caption{Summary of the performance of the \textit{semantic-only} mode of OmniLLP compared to other approaches (metrics are obtained from 5 bootstrap runs)}
\label{tab:full_semantic}
\end{table*}

%Additionally, semantic clustering significantly improves precision by approximately 7 percentage points (0.57 vs. 0.50), and reduces Average Ordinal Distance (AOD) by approximately half compared to the zero-shot approach. Notably, these gains hold across all studied projects, even those lacking clearly defined documentation-based components (e.g., Elasticsearch, Cassandra), further underscoring the effectiveness of our intent-based clustering method.

\begin{tcolorbox}[colback=gray!10, colframe=black, title=\textbf{Summary of RQ2}]
Despite labeling a large proportion of files as noise (up to 49\%), semantic-based retrieval statistically significantly enhances the predictive performance of LLM-LLPs, improving AUC by up to 6.5\% compared to documentation-based retrieval. This indicates that code semantic clusters --even with partial coverage-- offer a strong and practical value as a retrieval signal for LLM-based log level prediction.
\end{tcolorbox}

\subsection*{\textbf{RQ3. \RQIII}}
\textbf{Motivation}: Each of the signals studied earlier (i.e., semantic similarity and developer ownership) captures a different but complementary aspect of relationships between source files, each of which influences logging decisions. Individually, these signals have already demonstrated their value in improving the retrieval of relevant logging examples to guide LLM-LLPs. However, relying solely on one signal introduces key limitations that can hinder log level prediction. For instance, semantic clustering groups files purely based on functional similarity, thereby ensuring that the retrieved examples reflect code performing similar tasks; yet, it overlooks the fact that logging conventions often vary between teams, even for semantically similar code. Conversely, ownership clustering captures developer-specific logging practices by grouping files maintained by similar teams but ignores the possibility of similar logging intents existing across developer boundaries due to functional similarities. Given that logging decisions are influenced both by the functionality of code and by developer-specific logging strategies, we believe it is worthwhile to leverage these two signals simultaneously to retrieve ICL logging examples. In this RQ,  we investigate whether leveraging OmniLLP's multiplex clustering that takes into consideration the semantic and ownership signals simultaneously can improve the predictive performance of LLM-LLPs.

\textbf{Approach}: To quantify the impact of multiplex based retrieval on the performance of LLM-LLPs, we first perform the multiplex clustering for each studied project, as described in Section~\ref{sec:mlpx}. Then, we evaluate the performance of OmniLLP operating in "\textit{multiplex}" mode against three baselines: (a) random retrieval from the entire project, (b) OmniLLP's "\textit{ownership-only}" mode , and (c) OmniLLP's "\textit{semantic-only}" mode.

\begin{table*}[]
\resizebox{\textwidth}{!}{
\begin{tabular}{|l|lcc|lcc|ccc|ccc|}
\hline
\multicolumn{1}{|c|}{\textbf{Project}} & \multicolumn{3}{c|}{\textbf{\begin{tabular}[c]{@{}c@{}}Global \\ Few-shot (5)\end{tabular}}}       & \multicolumn{3}{c|}{\textbf{\begin{tabular}[c]{@{}c@{}}Ownership-based\\ Few-shot (5)\end{tabular}}} & \multicolumn{3}{c|}{\textbf{\begin{tabular}[c]{@{}c@{}}Semantic-based\\ Few-shot (5)\end{tabular}}} & \multicolumn{3}{c|}{\textbf{\begin{tabular}[c]{@{}c@{}}OmniLLP\\ Few-shot (5)\end{tabular}}}                         \\ \cline{2-13} 
\multicolumn{1}{|c|}{}                 & \multicolumn{1}{l|}{AUC}           & \multicolumn{1}{l|}{Precision}     & \multicolumn{1}{l|}{AOD} & \multicolumn{1}{l|}{AUC}            & \multicolumn{1}{l|}{Precision}      & \multicolumn{1}{l|}{AOD} & \multicolumn{1}{l|}{AUC}            & \multicolumn{1}{l|}{Precision}     & \multicolumn{1}{l|}{AOD} & \multicolumn{1}{l|}{AUC}                    & \multicolumn{1}{l|}{Precision}              & \multicolumn{1}{l|}{AOD} \\ \hline
Hadoop                                 & \multicolumn{1}{l|}{0.802 ± 0.004} & \multicolumn{1}{l|}{0.500 ± 0.008} & 0.806 ± 0.002            & \multicolumn{1}{l|}{0.802 ± 0.004}  & \multicolumn{1}{l|}{0.500 ± 0.008}  & 0.806 ± 0.002            & \multicolumn{1}{c|}{0.865 ± 0.002}  & \multicolumn{1}{c|}{0.587 ± 0.002} & 0.837 ± 0.003            & \multicolumn{1}{c|}{\textbf{0.914 ± 0.003}} & \multicolumn{1}{c|}{\textbf{0.681 ± 0.005}} & \textbf{0.876 ± 0.004}   \\ \hline
HBase                                  & \multicolumn{1}{c|}{0.805 ± 0.000} & \multicolumn{1}{c|}{0.502 ± 0.011} & 0.805 ± 0.003            & \multicolumn{1}{c|}{0.805 ± 0.000}  & \multicolumn{1}{c|}{0.502 ± 0.011}  & 0.805 ± 0.003            & \multicolumn{1}{c|}{0.863 ± 0.003}  & \multicolumn{1}{c|}{0.568 ± 0.007} & 0.838 ± 0.004            & \multicolumn{1}{c|}{\textbf{0.882 ± 0.002}} & \multicolumn{1}{c|}{\textbf{0.610 ± 0.008}} & \textbf{0.847 ± 0.004}   \\ \hline
Elasticsearch                          & \multicolumn{1}{c|}{0.844 ± 0.010} & \multicolumn{1}{c|}{0.496 ± 0.003} & 0.814 ± 0.001            & \multicolumn{1}{c|}{0.844 ± 0.010}  & \multicolumn{1}{c|}{0.496 ± 0.003}  & 0.814 ± 0.001            & \multicolumn{1}{c|}{0.924 ± 0.002}  & \multicolumn{1}{c|}{0.638 ± 0.008} & 0.863 ± 0.001            & \multicolumn{1}{c|}{\textbf{0.964 ± 0.001}} & \multicolumn{1}{c|}{\textbf{0.638 ± 0.008}} & \textbf{0.863 ± 0.001}   \\ \hline
Cassandra                              & \multicolumn{1}{l|}{0.825 ± 0.007} & \multicolumn{1}{c|}{0.481 ± 0.017} & 0.810 ± 0.008            & \multicolumn{1}{l|}{0.825 ± 0.007}  & \multicolumn{1}{c|}{0.481 ± 0.017}  & 0.810 ± 0.008            & \multicolumn{1}{c|}{0.863 ± 0.006}  & \multicolumn{1}{c|}{0.568 ± 0.012} & 0.841 ± 0.006            & \multicolumn{1}{c|}{\textbf{0.907 ± 0.004}} & \multicolumn{1}{c|}{\textbf{0.652 ± 0.020}} & \textbf{0.868 ± 0.005}   \\ \hline
\end{tabular}
}
\caption{Summary of the performance of the \textit{multiplex} mode of OmniLLP compared to other approaches (metrics are obtained from 5 bootstrap runs)}
\label{tab:full_omnillp}
\end{table*}

\textbf{Results}: \textbf{Multiplex retrieval significantly boosts the performance of LLM-LLPs, providing both improved accuracy and near-real-time retrieval speed across all studied projects}. In fact, leveraging OmniLLP to retrieve ICL logging examples from multiplex clusters results in a statistically significant (Wilcoxon test, $\alpha =0.01$, Cohen's d, $d>0.7$)  improvement in AUC ranging from 0.88 (observed for HBase) to 0.96 (observed for Elasticsearch), outperforming the global random retrieval baseline by 8\% to 10\% and the best single-layer (i.e., RQ1/RQ2) retrieval by 2\% to 4\%. Similarly, precision and AOD exhibit similar statistically significant (Wilcoxon test, $\alpha =0.01$, Cohen's d, $d>0.7$) improvements as the median Precision and AOD improves by 11\% to 18\% and 4\% to 7\% respectively. Furthermore, our retrieval step leverages FAISS~\footnote{https://github.com/facebookresearch/faiss} making it highly efficient, with a median latency of 8.9 ms per retrieval, effectively operating at real-time speed (i.e., suitable for IDEs). This ensures that the log level prediction bottleneck (if any) remains with the LLM's generation step rather than with the ICL example retrieval.

\textbf{Across our evaluated projects, most OmniLLP mispredictions consistently involve confusion between adjacent log levels and concentrate in a small subset of files}, as shown in Table~\ref{tab:multiplex_miss}. Specifically, we observe that adjacent-level mispredictions (e.g., confusing DEBUG with INFO, or WARN with ERROR) account for 64\% (Cassandra), to 72\% (Hadoop) of total mispredictions. Furthermore, a significant portion of mispredictions occurs in files that our multiplex clustering marked as noise (i.e., unclustered files), representing between 39\% (Elasticsearch) and 56\% (Cassandra) of total errors. Finally, mispredictions are heavily concentrated at the file level, with fewer than 17\% of files accounting for at least half of all mispredictions in each project. This pattern suggests that predictive difficulties faced by OmniLLP are highly localized, providing clear opportunities for targeted improvements.

%Specifically, between 64.1% (Cassandra) and 72.8% (Hadoop) of mispredictions involve adjacent log-level pairs (e.g., predicting DEBUG instead of INFO). Additionally, between 39.4% (Elasticsearch) and 56.4% (Cassandra) of mispredictions occur in files that were not strongly associated with any multiplex cluster (noise cluster -1). Finally, mispredictions are heavily concentrated at the file level, with fewer than 17% of files accounting for at least half of all mispredictions in each project. This pattern suggests that predictive difficulties are highly localized, providing clear opportunities for targeted improvements in OmniLLP's predictions.

\begin{table}[]
\resizebox{0.5\textwidth}{!}{
\begin{tabular}{|l|l|l|l|}
\hline
\multicolumn{1}{|c|}{\textbf{Project}} & \multicolumn{1}{c|}{\textbf{\begin{tabular}[c]{@{}c@{}}Adjacent log level\\  Confusion (\%)\end{tabular}}} & \multicolumn{1}{c|}{\textbf{Noise Cluster (\%)}} & \multicolumn{1}{c|}{\textbf{\begin{tabular}[c]{@{}c@{}}Files Covering 50\% \\ of Mispredictions (\%)\end{tabular}}} \\ \hline
\textbf{Hadoop}                        & 72.8\%                                                                                                     & 41.2\%                                           & 14.7\%                                                                                                              \\ \hline
\textbf{Cassandra}                     & 64.1\%                                                                                                     & 56.4\%                                           & 13.3\%                                                                                                              \\ \hline
\textbf{HBase}                         & 67.0\%                                                                                                     & 54.5\%                                           & 13.9\%                                                                                                              \\ \hline
\textbf{Elasticsearch}                 & 64.7\%                                                                                                     & 39.4\%                                           & 16.5\%                                                                                                              \\ \hline
\end{tabular}}
\caption{Overview of the misspredictions using OmniLLP (multiplex mode)}
\label{tab:multiplex_miss}
\end{table}

\begin{tcolorbox}[colback=gray!10, colframe=black, title=\textbf{Summary of RQ3}]
Multiplex clustering, which simultaneously integrates semantic and ownership signals, provides a retrieval strategy that significantly boosts the AUC of LLM-based LLPs (up to 12\% improvement over random retrieval). We recommend using OmniLLP’s multiplex mode for real-time retrieval of high-quality ICL examples supporting log level prediction.
\end{tcolorbox}

\section{Threats to validity}

\subsection{Internal Validity}
An internal validity threat concerns our reliance on embedding models to represent files for clustering and retrieval. While we employed a state-of-the-art embedding model (CodeXEmbed) due to its proven effectiveness in capturing code semantics, any inherent biases or limitations in this model could influence the quality of our clustering outcomes, consequently affecting the performance of LLM-LLPs using those clusters. To mitigate this threat, we validated cluster quality using established metrics (Silhouette, DBCV, and modularity) and assessed their stability through bootstrapping and temporal re-clustering analyses. Future work could benefit from exploring different embedding approaches to further strengthen the generalizability and robustness of the results.

\subsection{Construct Validity}
A construct validity threat involves the selection of clustering hyperparameters, such as the resolution parameter ($\gamma$) and cluster size thresholds (e.g., minimum cluster size). Arbitrary or suboptimal parameter choices might impact clustering results. To address this concern, we conducted specialized grid searches for each of our projects, identifying optimal parameter values based on robust clustering quality evaluation metrics, therefore ensuring our results' consistency and reducing sensitivity to hyperparameter choices.

\subsection{External Validity}
Regarding external validity, our empirical evaluation focused on four large-scale, widely-used open-source Java projects (Hadoop, HBase, Elasticsearch, and Cassandra). Although these projects represent significant open-source software systems, our findings might not fully generalize to smaller, proprietary, or non-Java software projects. Future work might replicate our analysis across diverse software ecosystems and languages. 
Another external validity threat relates to our reliance on a single large language model (CodeLlama-7B) for log level prediction. We deliberately selected CodeLlama-7B because it strikes an ideal balance between predictive performance and practical usability, aligning closely with our OmniLLP design goals, which emphasize efficiency and deployment feasibility. Nonetheless, the observed results may still be influenced by model-specific behaviors, and different LLM architectures might yield varying outcomes. To mitigate this concern, we leveraged multiple robust evaluation metrics (AUC, precision, AOD) and conducted statistical analyses of performance improvements. Future studies might explore different LLM architectures and their impact on log level prediction.

One final external validity threat relates to OmniLLP's reliance on two specific clustering signals (semantic similarity and developer ownership). While these signals have demonstrated significant predictive performance improvements, other signals (e.g., defect density) might also influence developers' log level decisions and could lead to further performance gains. Nevertheless, our results show that our selected signals represent two core dimensions reflecting real-world logging practices (i.e., shared functionality and authorship), thus supporting the validity and utility of our hypothesis. Future research could explore incorporating additional signals to further enhance OmniLLP's predictive capabilities.

\section{Conclusion}
Selecting appropriate log levels for new logging statements is an important, yet challenging software engineering task. Prior studies have leveraged machine learning models including large language models (LLMs) to automate log level predictions. However, these approaches often overlook the inherent multi-component structure of modern software systems, treating projects as monolithic entities and thus missing opportunities to leverage internal structural information to enhance the prediction performance of LLPs.

In this paper, we introduced OmniLLP, a retrieval-augmented generation framework that significantly enhances log level prediction by simultaneously integrating semantic (i.e., what the code does) and ownership (i.e., who wrote the code) information through multiplex clustering. Our evaluation across four popular open-source projects (Hadoop, HBase, Elasticsearch, and Cassandra) demonstrates that multiplex clustering effectively yields compact, and temporally stable clusters. Leveraging these clusters for ICL retrieval leads to LLPs that outperform single-signal (e.g., ownership) LLP baselines and achieves impressive predictive accuracy (AUC improvement between 8\% and 12\% compared to prior work).

Our findings underscore the value of considering relevant software engineering knowledge (such as semantic intent and developer ownership) for automated log level prediction. We envision OmniLLP being practically deployed within IDEs, providing real-time recommendations to developers, thereby enhancing logging practices and software maintainability. Beyond log level prediction, we plan to extend OmniLLP to support additional logging tasks, such as log message generation and log placement recommendations, further assisting developers in producing effective, informative, and well-structured logs.

\bibliographystyle{ACM-Reference-Format}
\bibliography{sample-base}

\end{document}